\documentclass[usenatbib,letterpaper]{mn2e}					
\usepackage{graphicx}
\usepackage{amsmath}
\usepackage{amssymb}
\title[Cross-Correlation of CFHTLenS Galaxy Number Density and Planck CMB Lensing]{Cross-Correlation of CFHTLenS Galaxy Number Density and Planck CMB Lensing}
\author[Y. Omori and G. Holder]{Y. Omori$^{1}$\thanks{E-mail:yomori@physics.mcgill.ca} and G. Holder$^{1}$\\
$^{1}$McGill University, QC, Canada, H3A2T8}

\begin{document}


\pagerange{\pageref{firstpage}--\pageref{lastpage}} \pubyear{2002}

\maketitle

\label{firstpage}

\begin{abstract}
We measure the cross-power spectrum between galaxy density from Canada-France-Hawaii-Telescope Lensing Survey (CFHTLenS) catalogues and gravitational lensing convergence from Planck data release 1 (2013) and 2 (2015). We investigate three main galaxy samples: $18.0<i_{\rm AB}<22.0$, $18.0<i_{\rm AB}<23.0$, $18.0<i_{\rm AB}<24.0$ in the redshift range $0.2<z<1.3$ in each of the four CFHTLenS wide fields. By comparing the measured cross-spectrum with model predictions, linear galaxy-dark matter biases of  $b=0.82^{+0.24}_{-0.23}, 0.83^{+0.19}_{-0.18}, 0.82^{+0.16}_{-0.14}$ are inferred at significances of $3.5, 4.5, 5.6\sigma$ using the Planck 2015 release. These measurements are marginally consistent with biases derived from galaxy-galaxy auto-correlations: $b=1.15^{+0.02}_{-0.01}, 1.08^{+0.01}_{-0.01}$ and $0.96^{+0.01}_{-0.01}$ respectively. Using the 2013 Planck release, we obtain biases of  $b=1.33^{+0.29}_{-0.28}, 1.19^{+0.23}_{-0.23}, 1.16^{+0.19}_{-0.18}$, showing significant differences between the releases.

\end{abstract}


\section{Introduction}
An approach to measuring the distribution of dark matter in our Universe is by using weak gravitational lensing, where light is deflected by mass between the source and the observer. In the case of cosmic microwave background (CMB) weak lensing, the paths of photons emitted from the last scattering surface become perturbed by the large-scale structure in the Universe amidst their journey to the observer. This changes the statistics of hot/cold spots size distribution, leading to a change in the local angular power spectrum on the scale of the primary acoustic peaks. \citep{lewis06}. Since the degree of deflection is proportional to the integrated potential that the photon traverses, weak lensing is a sensitive probe for studying the large-scale structure between the CMB and us.

The clustering properties and statistics of dark matter halos have been extensively studied through simulations \citep{millennium09, klypin11}. Since these dark matter halos form in peaks in the peaks of the background density distribution, they are clustered more strongly than the background matter distribution~\citep{kaiser84,bardeen86}.  Although galaxies live inside dark matter halos, the formation processes and their dynamics make the distribution of galaxies non-local and possibly stochastic relative to the halo center, making the clustering characteristics different~\citep{mo}. The ratio between the clustering amplitude of dark matter and galaxies is called linear galaxy bias.

Here we cross-correlate the galaxy number density from the Canada-France-Hawaii Telescope Lensing Survey (CFHTLenS) \citep{heymans12,erben13,velander14} and the all-sky convergence map from Planck (Planck XV 2015, hereafter PXV2015) to measure galaxy biases. Similar studies include Planck lensing cross-correlated with NRAO VLA Sky Survey (NVSS) galaxies, MaxBCG clusters, SDSS LRGs and WISE (Planck XVII 2013, hereafter PXVII2013), Wilkinson Microwave Anisotropy Probe (WMAP) lensing cross-correlated with NVSS galaxies \citep{smith07}, WMAP lensing cross-correlated with  LRGs and quasar from SDSS~\citep{hirata08} Blanco Cosmology Survey (BCS) galaxies cross-correlated with South Pole Telescope (SPT) lensing~\citep{bleem12}, Sloan Digital Sky Survey (SDSS) quasar maps cross-correlated with Atacama Cosmology Telescope (ACT) lensing~\citep{sherwin12}, Wide-field Infrared Survey Explorer (WISE) selected quasars cross-correlated with SPT lensing~\citep{geach13}, Herschel/SPIRE cosmic infrared background (CIB) cross-correlated with SPT lensing~\citep{holder13}, SPTpol E-mode polarization cross-correlated with estimates of the lensing potential from Herschel/SPIRE CIB~\citep{hanson13}, Polarbear experiment CMB polarization cross-correlated with Herschel/SPIRE CIB~\citep{polarbear} and reconstructed convergence from ACTpol cross-correlated with CIB measurements from Planck~\citep{vanengelen14b}.  In our study, we  present both the cross-correlations between galaxy overdensity and lensing, and the galaxy auto-correlations using the \emph{same} galaxy data, and compare the linear galaxy biases that we obtain.

In this paper we assume flat $\Lambda$CDM cosmology $(\Omega_{m}\hspace{-0.1cm}=\hspace{-0.1cm}0.28,\ \Omega_{\Lambda}\hspace{-0.08cm}=\hspace{-0.08cm}0.72,\ H_{0}\hspace{-0.08cm}=\hspace{-0.08cm}100 h\ {\rm km\ s}^{-1} {\rm Mpc}^{-1} {\rm\ and\ } \sigma_{8}\hspace{-0.08cm}=\hspace{-0.08cm}0.82,\ n_{s}\hspace{-0.08cm}=\hspace{-0.08cm}0.96)$ with $h\hspace{-0.08cm}=\hspace{-0.08cm}0.70$. All magnitudes are given in the AB system.

\section{Theory}
CMB lensing convergence is calculated by integrating the matter fluctuation in the line-of-sight direction:
\begin{equation}
\kappa({\bf \hat{n}})=\int d\chi W^{\kappa}(\chi)\delta[\chi\hat{\rm \bf n},z(\chi)],
\end{equation}
where ${\bf \hat{n}}$ is the line-of-sight direction, $\chi$ is the line-of-sight comoving distance, $\delta$ is the fractional dark matter density fluctuation and $W^{\kappa}$ is the lensing kernel: 
\begin{equation}\label{eq:k_kernel}
W^{\kappa}(\chi)=\frac{3}{2}\Omega_{{\rm m},0}\left(\frac{H_{0}}{c}\right)^{2}\frac{\chi}{a(\chi)}\frac{\chi_{\rm CMB}-\chi}{\chi_{\rm CMB}},
\end{equation}
which describes the efficiency of lensing at a given redshift when multiplied by $d \chi/dz$.

\begin{figure}
\begin{center}
\hspace{-0.2cm}\includegraphics[width=0.085\textwidth,bb= 175 -5 250 470]{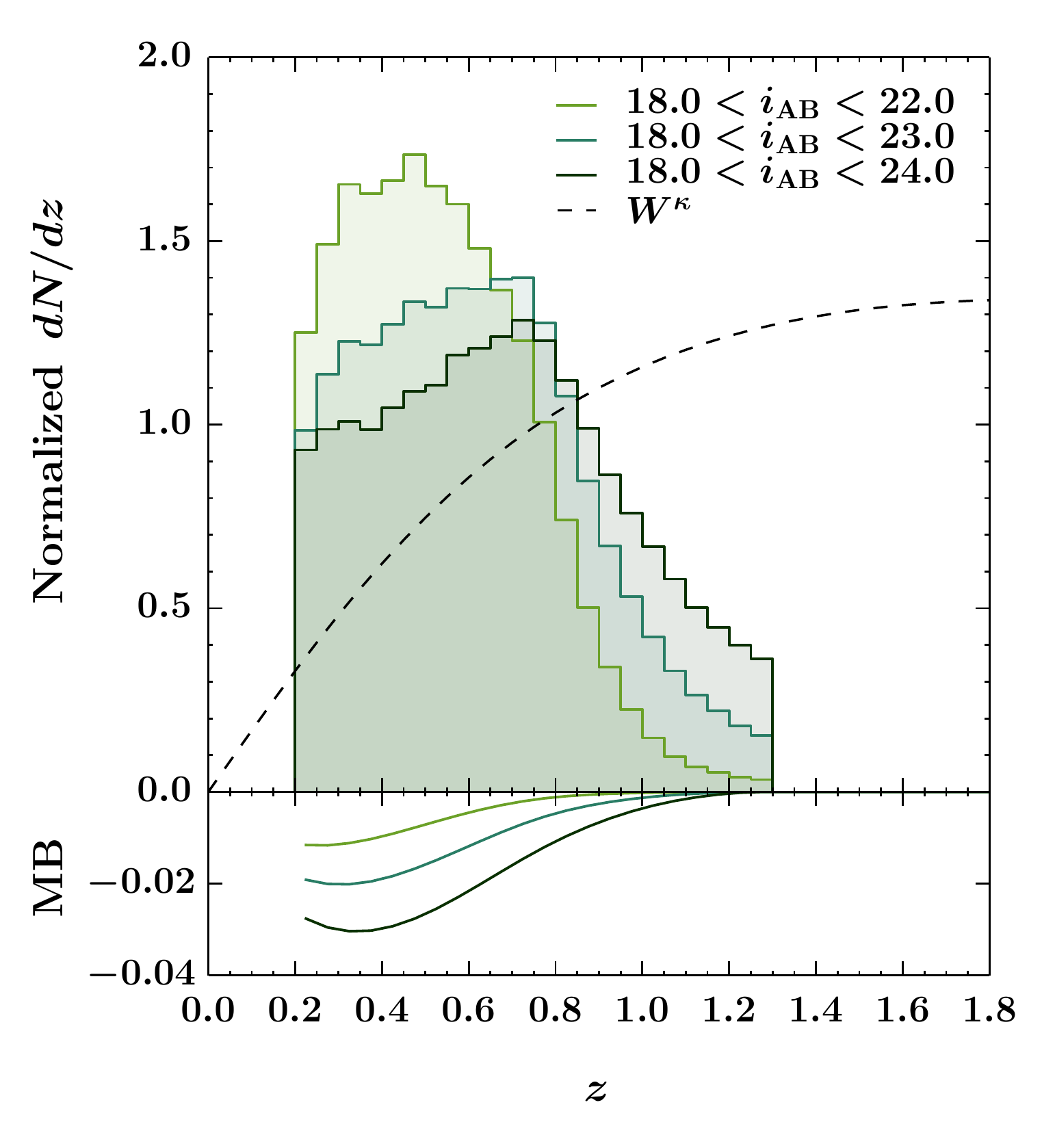}
\caption{{\bf Light green, Turquoise, Dark green:} Redshift distribution $dn/dz$ for the $18.0<i_{\rm AB}<22.0$, $18.0<i_{\rm AB}<23.0$, $18.0<i_{\rm AB}<24.0$ samples in the range $0.2<z<1.3$. The distribution is normalized such that the integral $\int d z(d N/d z)= 1$. {\bf Dashed line:} Convergence kernel $W^{\kappa}$ multiplied with ($d\chi/d z$) and arbitrary normalization applied for visualization purposes to demonstrate the effectiveness of the lenses at various redshifts. {\bf Lower panel:} Amplitude of magnification bias for each of the galaxy samples.}
\label{fig:dndz}
\end{center}
\end{figure}

Under the assumption of the galaxy distribution tracing out the peaks of dark matter density fluctuations, the overdensity of galaxies can be described as: 
\begin{equation}
g({\bf \hat{n}})=\int d \chi W^{g} (\chi)\delta[\chi\hat{\rm \bf n},z(\chi)].
\end{equation}
The distance kernel $W^{g}$\citep{sherwin12}:
\begin{equation}\label{eq:g_kernel}
W^{g}(\chi)=\frac{1}{\left[\int d z'\frac{d N(z')}{dz'}\right]}\frac{d z}{d \chi}\frac{d N(z)}{d z}b(\chi)+MB(\chi)\\
\end{equation}
describes the distribution of galaxies as a function of redshift in terms of co-moving distances. The second term in equation \ref{eq:g_kernel} is the magnification bias, where the observed number density of galaxies are altered due to lensing caused by mass between the foreground galaxy and the observer \citep{turner84} is given by:
\begin{equation}
MB(\chi)=\frac{3}{2}\Omega_{{\rm m},0}\frac{c}{H(z(\chi))}\left(\frac{H_{0}}{c}\right)^{2}(1+z(\chi))g(\chi)\left(5s-2\right),
\end{equation}
with
\begin{equation}
g(\chi)=\chi \int_{\chi}^{\chi_{\rm CMB}} d\chi' \frac{\chi'-\chi}{\chi'}\frac{dN/dz}{\int dz'\ dN/dz'}\frac{dz}{d\chi'},
\end{equation}
and
\begin{equation}
s=\frac{d{\rm log}_{10}N(<m)}{dm}\Big|_{m_{\rm lim}},
\end{equation}
for a galaxy sample with limiting magnitude $m_{\rm lim}$ \citep{villumsen95}. As shown in figure \ref{fig:dndz}, the amplitudes of the magnification biases for each sample are significantly smaller than the $dN/dz$ term.   

The cross-spectrum between convergence and galaxy overdensity under the Limber approximation is the power spectrum weighted by the lensing and galaxy kernels~ \citep{limber53,kaiser92}:
\begin{equation}
C_{\ell}^{\kappa g}=\int dz \frac{d \chi }{d z}\frac{1}{\chi^{2}}W^{\kappa}(\chi)W^{g}(\chi)P\left(k=\frac{\ell}{\chi},z\right),
\label{eq:clkg}
\end{equation}
where $P(k,z)$ is the non-linear matter power spectrum at redshift $z$. As a comparison, we also compute the galaxy overdensity auto-spectrum given by:
\begin{equation}
C_{\ell}^{gg}=\int d z \frac{d \chi }{ d z}\frac{1}{\chi^{2}}W^{g}(\chi)^{2}P\left(k=\frac{\ell}{\chi},z\right).
\label{eq:clgg}
\end{equation}
\vspace{0.1cm}
The model spectra in equations \ref{eq:clkg}, \ref{eq:clgg}  are computed using the non-linear power spectrum from \texttt{CAMB} \citep{camb} and revised \texttt{Halofit}~\citep{smith03,takahashi12}. The redshift distributions $dN/dz$ for each subsample are produced by averaging the individual photometric redshift probability distributions for each galaxy rather than simply using the best-fit estimates for each galaxy.

\section{Galaxy maps}
In this study, we use the galaxies from the CFHTLenS\footnote{http://cfhtlens.org/} galaxy survey. CFHTLenS is part of the wide component of the Canada-France-Hawaii Telescope Legacy Survey \footnote{http://www.cfht.hawaii.edu/Science/CFHTLS/} (CFHTLS), which consists of 4 fields centered at 2:18:00 -07:00:00, 08:57:49 -03:19:00, 14:17:54 +54:30:31, 22:13:18 +01:19:00, each ranging from $23-64\ {\rm deg}^{2}$ with a total of $154\ {\rm deg}^{2}$ and full multi-colour depth of $i_{\rm AB}=24.7$ \citep{heymans12,erben13,velander14}. We limit our analysis to galaxies in the redshift range $0.2<z<1.3$, which is confirmed by \cite{heymans12} to have a photometric redshift distribution that resembles the measured spectroscopic redshift distribution. Galaxies selected with $i_{\rm AB}<24.5$ in this redshift slice have a scatter of $0.03<\sigma_{\rm \Delta z}< 0.06$ (where $\sigma_{\rm \Delta z}^{2}$ is the variance in the difference between the photometric and spectroscopic redshifts $(z_{\rm p}-z_{\rm s})/(1+z_{\rm s}))$ with $10\%$ of the galaxies classified as outliers \citep{hildebrandt12,benjamin13}. Within this redshift slice, three samples with $18.0<i_{\rm AB}<22.0$, $18.0<i_{\rm AB}<23.0$, $18.0<i_{\rm AB}<24.0$ are extracted, resulting  in galaxy catalogues with mean number densities $n=3.4, 7.7, 15.4$ per square arcminute respectively. The lower magnitude cut is applied to reduce dispersion and bias due to photometric redshift estimates in addition to removal of possible stars that passed through initial reduction pipeline  classifications (\texttt{star\_flag}\ $=0$ and \texttt{mask\ }$=0$, descriptions of each flag can be found in \cite{erben13}).  

Using these catalogues, the fractional galaxy overdensity maps at arcminute resolution are produced by:
\begin{equation}
\delta_{ij}=\frac{N_{ij}-\langle N\rangle w_{ij}}{\langle N\rangle w_{ij}},
\end{equation}
for each $i$-th and $j$-th pixel, where $w_{ij}$ is the pixel-by-pixel weight map produced by degrading the arcsecond resolution masks to match the same resolution and 
\begin{equation}
\langle N \rangle = \frac{\sum_{ij} N_{ij}}{\sum_{ij} w_{ij}}.
\end{equation}

\section{Convergence maps}
We use the observed and simulated convergence maps from the Planck 2015 release\footnote{http://irsa.ipac.caltech.edu/data/Planck/release\_2/all-sky-\\maps/maps/component-maps/lensing/COM\_CompMap\_Lensing\_\\2048\_R2.00.tar} (PXV2015) and compare the results with the 2013 release\footnote{http://irsa.ipac.caltech.edu/data/Planck/release\_1/all-sky-\\maps/previews/COM\_CompMap\_Lensing\_2048\_R1.10/index.html} (PXVII2013). The 2015 convergence map is produced by applying a quadratic estimator~\citep{okamoto03} on a SMICA component separated map, which is synthesized by cleaning the foregrounds and combining all nine frequency bands. The Galaxy and point sources are masked in the process and the bands are restricted in the range of $100\leq\ell\leq2048$ (PXV2015). The 2013 map is released as a ${\bar\phi}$ map is generated by combining the 143 and 217 GHz channels with dust distribution subtracted using the 857 GHz channel. The map also masks point sources included in the ERCSC, SZ and PCCS catalogs (PXVII2013).   The map is transformed into a convergence $\kappa$ map by taking the transform:
\begin{equation}
\kappa(\boldsymbol{\ell})=\frac{\ell(\ell+1)}{2}(\mathcal{R}_{\ell}^{\phi})^{-1}{\bar\phi}(\boldsymbol{ \ell}),
\end{equation}
where $\mathcal{R}_{\ell}^{\phi}$ is a normalization factor. We combine the Planck 2013 and 2015 masks and apply this to both convergence maps for consistency. Additional masking is not applied beyond the masks released by the Planck collaboration. Field regions were extracted by multiplying this all-sky mask with galaxy catalogue masks.

Biases in the lensing reconstruction due to the thermal Sunyaev-Zel’dovich (tSZ) effect or cosmic infra-red background (CIB) fluctuations are expected to be small with Planck's angular resolution~\citep{vanengelen14a,osborne14}.

\section{Cross and auto-spectrum}

The cross-power spectrum is calculated using the flat-sky approximation and multiplying the maps in Fourier space:
\begin{equation}
C_{\ell}^{\kappa g} =f\langle Re(({\mathcal K({\bf l}) - \langle\mathcal K_{\rm sim}({\bf l})\rangle})^{*}{\mathcal G({\bf l})}) \rangle \rvert_{\bf l \in \ell}
\end{equation}
where $f$ is the normalization factor $N_{1}N_{2}/\sum{w_{ij}}$ and ${\mathcal K}$, ${\mathcal G}$ are the Fourier transforms of the convergence and the galaxy overdensity maps multiplied with the weight map. The auto-spectrum is calculated in a similar way but replacing ${\mathcal G}$ with ${\mathcal K}$ and subtracting the shot noise contribution $C_{\ell}^{\rm shot}=1/\langle n\rangle$, where $\langle n\rangle$ is the average number density of galaxies per steradian.

Uncertainties are calculated by cross-correlating the galaxy maps with 100 simulated Planck convergence maps (released by the Planck collaboration) reduced  through the same pipeline, and calculating the variance for each bin. Using these uncertainties, the weighted averages for each individual bin are calculated.

\begin{figure}
\begin{center}
\hspace{-0.3cm}\includegraphics[width=0.078\textwidth,bb= 175 -5 250 425]{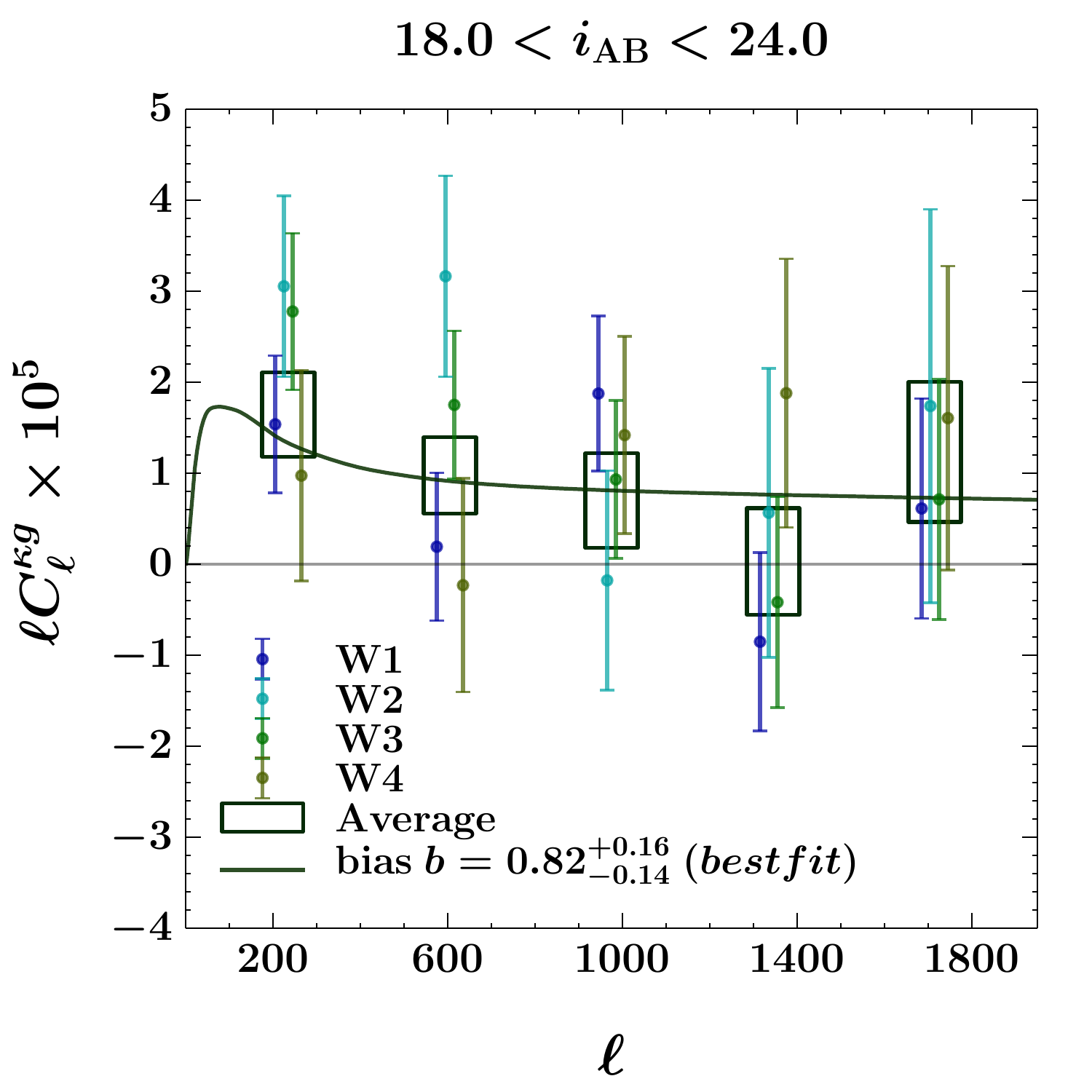}
\caption{Cross-correlation of $18.0<i_{\rm AB}<24.0$ sample for W1,W2,W3,W4, and the average of the four fields with the best fit theory model.}
\label{fig:4fields}
\end{center}
\end{figure}

Since the uncertainties in the galaxy auto-spectrum cannot be derived in the same way, we resort to an internal block-jackknife method often used in 2-point correlation function analysis, where we divide each of the fields into 64 sub-regions and the jackknife samples are then created by omitting each of these sub-samples \citep{zehavi05,norberg09}. The covariance matrix is estimated by:

\begin{equation}
{\Sigma}^{gg}=\frac{N_{\rm sub}-1}{N_{\rm sub}}\sum_{n=1}^{N_{\rm sub}}(C_{\ell}^{n}-\langle C_{\ell}\rangle)_{i}(C_{\ell}^{n}-\langle C_{\ell}\rangle)_{j},
\end{equation}
while the covariance matrix for the cross-correlation is calculated by:
\begin{equation}
{\Sigma}^{\kappa g}=\frac{1}{N_{\rm maps}}\sum_{m=1}^{N_{\rm maps}}(C_{\ell}^{m}-\langle C_{\ell}\rangle)_{i}(C_{\ell}^{m}-\langle C_{\ell}\rangle)_{j}.
\end{equation}
Using these covariance matricies, the best fit biases are derived by minimizing the $\chi^{2}$:
\begin{equation}
\chi^{2}=\sum_{ij}(C_{\ell-th}^{X}-C_{\ell}^{X})_{i}({\Sigma}^{X})_{ij}^{-1}(C_{\ell-th}^{X}-C_{\ell}^{X})_{j},
\end{equation}
where $X=\kappa g$ or $gg$ and $i,j$ are the bin numbers. The significance of detection is defined by $\Delta \chi^{2}=\chi^{2}(b=0)-\chi^{2}_{\rm min}(b)$.

\section{Results}

\begin{figure*}
\begin{center}
\includegraphics[width=1.00\textwidth]{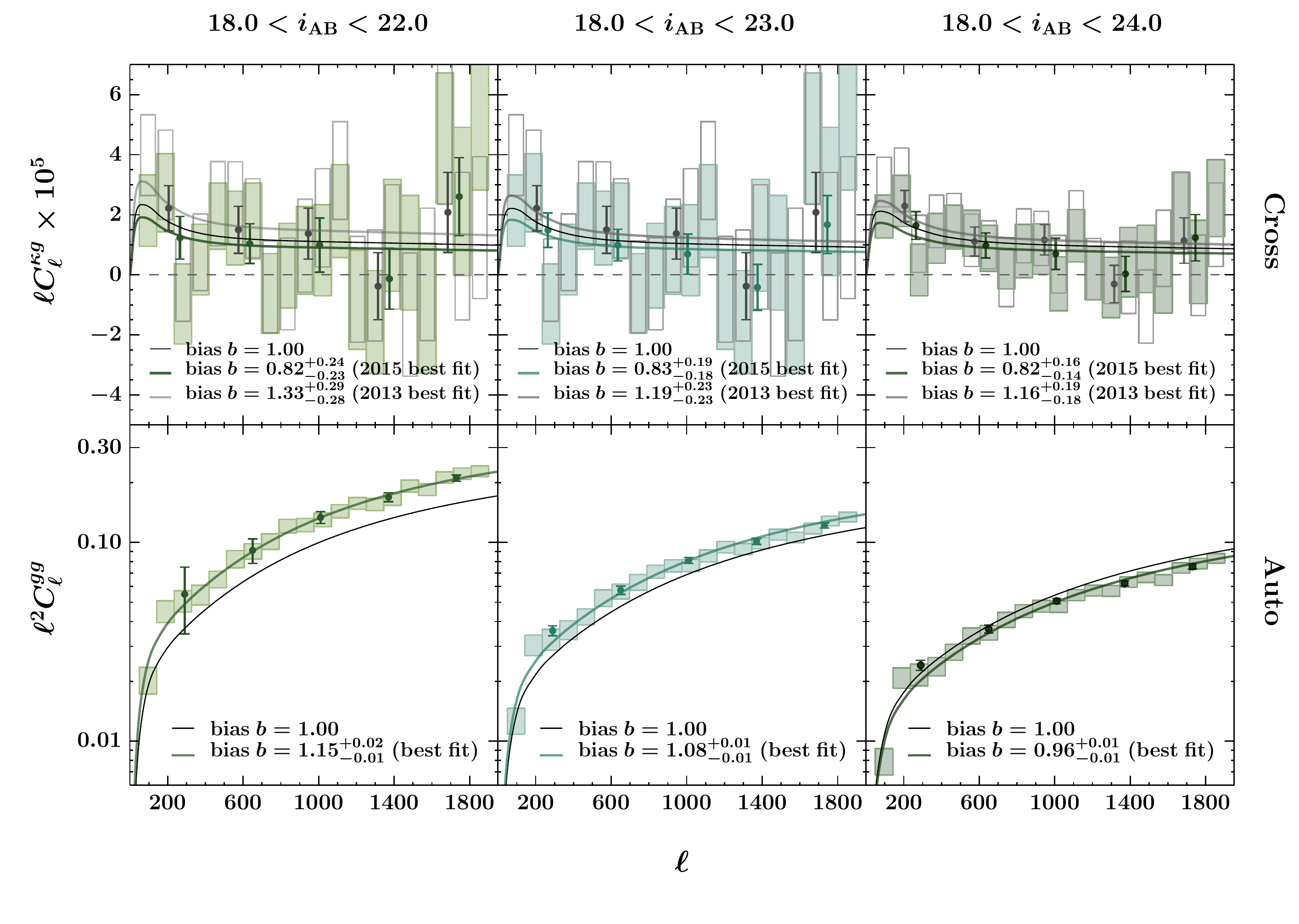}\\
\caption{Cross-spectrum $C_{\ell}^{\kappa g}$ for the $18.0<i_{\rm AB}<22.0$, $18.0<i_{\rm AB}<23.0$, $18.0<i_{\rm AB}<24.0$ galaxy samples cross-correlated with Planck convergence (filled bars: 2015, empty bars: 2013) (upper row) and the auto-spectrum using the same galaxy samples (lower row). The points with errorbars shown are the coarsely binned calculations for displaying purposes, and the $\chi^{2}$ are the fits to the 20 bins shown in the background.}
\label{fig:clkg}
\end{center}
\end{figure*}

\begin{figure*}
\begin{center}
\includegraphics[width=0.85\textwidth]{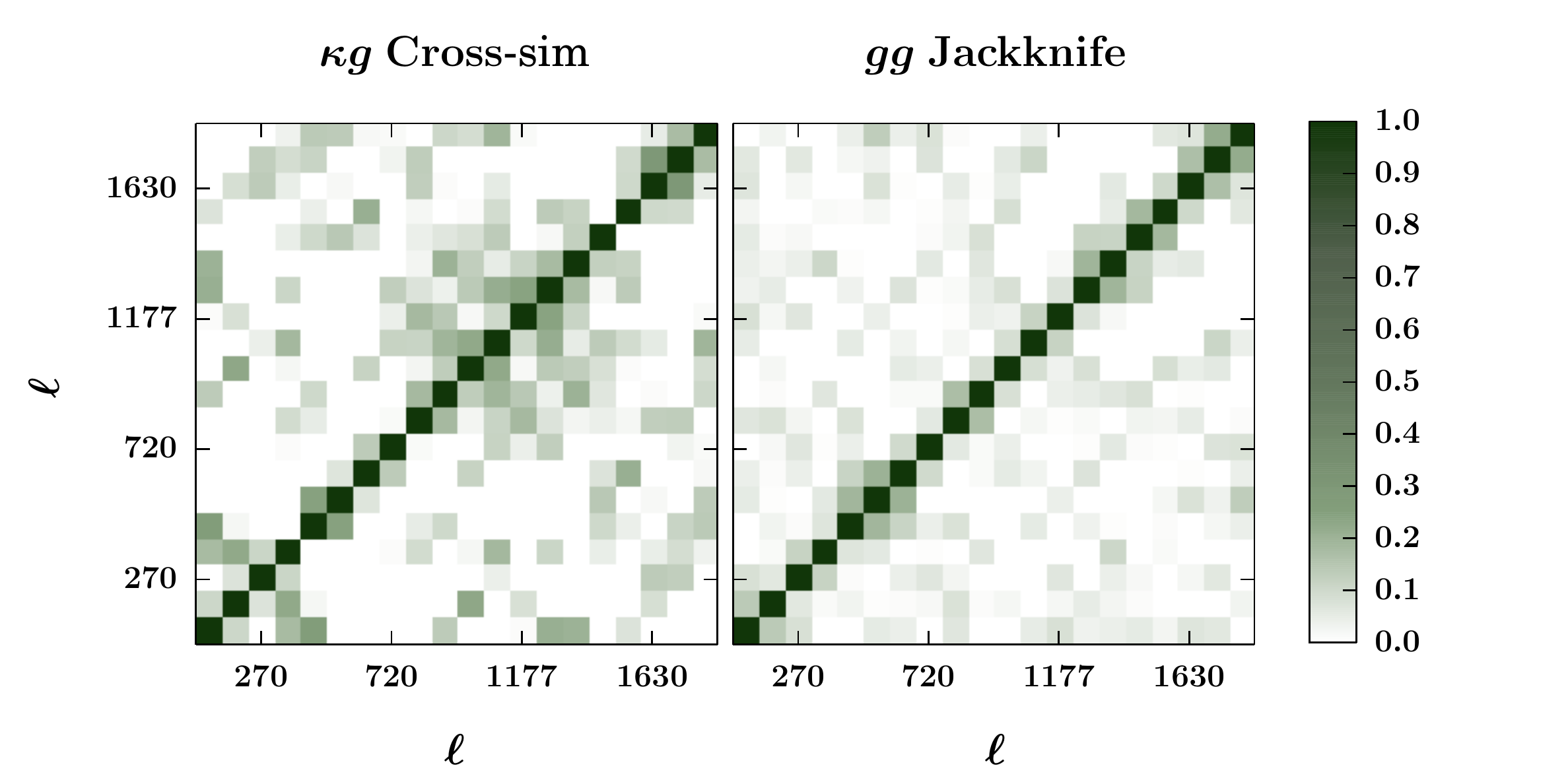}
\caption{Correlation matrices between bins for cross-correlation and auto-correlation.}
\label{fig:covar}
\end{center}
\end{figure*}

The cross-spectra for each of the fields using the $18.0<i_{\rm AB}<24.0$ sample are shown in figure \ref{fig:4fields}, and the different magnitude samples are shown in figure \ref{fig:clkg}. The best fit values $b$ and their associated $\chi^{2}$ values are listed in table \ref{table:bias}, where the $\chi^{2}$ values were calculated using 20 bins in the range of $\ell=50-1900$.

The best fit biases for the Planck 2015 release are found to be $b=0.82^{+0.24}_{-0.23}, 0.83^{+0.19}_{-0.18}, 0.82^{+0.16}_{-0.14}$ at significances of $3.5, 4.5, 5.6\sigma$ respectively. The  minimum $\chi^{2}$ for the fits  are 25.4, 18.6, 15.0 with a probability-to-exceed (PTE) of 0.15, 0.48, and 0.72. These best fit biases are significantly lower than the biases found from the Planck 2013 release:
$b=1.33^{+0.29}_{-0.28}, 1.19^{+0.23}_{-0.23}, 1.16^{+0.19}_{-0.18}$. Similarly, the best fit biases from $C_{\ell}^{gg}$ are found to be $1.15^{+0.02}_{-0.01}, 1.08^{+0.01}_{-0.01}, 0.96^{+0.01}_{-0.01}$, with $\chi^{2}$ of 14.1, 22.1, 38.0. Sample correlation matricies are shown in figure \ref{fig:covar}.

\begin{table*}
\begin{center}
\begin{tabular}{cccccccccccc}
\hline\\[-0.25cm]
\vspace{0.05cm}
Sample		& $N$ & $\langle n\rangle\ [{\rm arcmin}^{-2}]$ &\hspace{0.5cm} & $b\ (C_{\ell}^{gg})$ & $\chi^{2}$ & \hspace{0.5cm} &$b_{2015}\ (C_{\ell}^{\kappa g})$ & $\chi^{2}$ & \hspace{0.5cm} &$b_{2013}\ (C_{\ell}^{\kappa g})$ & $\chi^{2}$\\[-0.00cm] \hline\\[-0.2cm]
$18<i_{\rm AB}<22$ & 1444906 & 3.4  & & $1.15^{+0.02}_{-0.01}$ & 14.1 & & $0.82^{+0.24}_{-0.23}$ & 25.4 & & $1.33^{+0.29}_{-0.28}$ &15.4\\
$18<i_{\rm AB}<23$ & 3268383 & 7.7  & & $1.08^{+0.01}_{-0.01}$ & 22.1 & & $0.83^{+0.19}_{-0.18}$ & 18.6 & & $1.19^{+0.23}_{-0.23}$ &15.9\\
$18<i_{\rm AB}<24$ & 6529185 & 15.4 & & $0.96^{+0.01}_{-0.01}$ & 38.0 & & $0.82^{+0.16}_{-0.14}$ & 15.0 & & $1.16^{+0.19}_{-0.18}$ &15.4\\[0.1cm]\hline\\[-0.2cm]
\end{tabular}
\caption{\vspace{0.2cm}Table of best fit linear bias and $\chi^{2}$ values calculated using 20 bins between $50<\ell<1900$.}
\label{table:bias}
\end{center}
\end{table*}
These results suggest that unlike quasars, which are strongly biased tracers of dark matter (see Geach et al. 2012, Sherwin et al. 2012), optically-selected galaxies in these magnitude ranges on average are \emph{unbiased} tracers of dark matter fluctuations.

\section{Conclusion}

We have presented the results of cross-correlations between galaxy number density from CFHTLenS and CMB lensing convergence from Planck, along with galaxy-galaxy auto-correlations. We obtain the dark matter-galaxy bias using these two \emph{independent} methods. For the cross-correlations of the 3 magnitude samples, we obtain best fit biases of $b=0.82^{+0.24}_{-0.23}, 0.83^{+0.19}_{-0.18}, 0.82^{+0.16}_{-0.14}$ in the range of $50<\ell<1900$ using the 2015 release and $b=1.33^{+0.29}_{-0.28}, 1.19^{+0.23}_{-0.23}, 1.16^{+0.19}_{-0.18}$ using the 2013 release. From the galaxy auto-correlations, we obtain  biases of $b=1.15^{+0.02}_{-0.01}, 1.08^{+0.01}_{-0.01}, 0.96^{+0.01}_{-0.01}$, which settle between the values obtained from the two data releases and agrees well with previously established results that indicate linear galaxy bias of $\simeq 1$ at large scales (see for example Gaztanaga \& Frieman 1994). 

Although the galaxy-galaxy auto-correlations place stronger constraints on the value of the bias, these cross-correlations provide a more direct measurement of the relationship between galaxies and two-dimensional projected mass, because the  degree of photon path deflection is directly correlated with the depths of foreground gravitational potentials that it passes through.  Furthermore, cross-correlations are less sensitive to the complexity of survey masks and any hidden systematic errors that could be present in any single survey.

The power of cross-correlations is yet to be used at full potential. With the advent of upcoming wide galaxy surveys probing to fainter magnitudes, the sky coverage will improve the statistics while the depth will provide sufficient galaxy number density to maintain signal above shot-noise. For example, similar cross-correlation analyses will be performed between the Dark Energy Survey (DES) galaxies - SPT/SPTpol lensing and Hyper Supreme Camera (HSC) survey galaxies - ACT/ACTpol lensing, which will have several thousand square degrees of overlap.  With such large areas and fainter magnitude limits, the signal to noise is expected to be an order of magnitude larger. 

\section{Acknowledgements}
The authors thank Ludovic Van Waerbeke, Joachim Harnois-D\'eraps and Duncan Hanson for helpful discussions and valuable comments on the manuscript. This work is based on observations obtained with MegaPrime/MegaCam, a joint project of CFHT and CEA/DAPNIA, at the Canada-France-Hawaii Telescope (CFHT) which is operated by the National Research Council (NRC) of Canada, the Institut National des Sciences de l'Univers of the Centre National de la Recherche Scientifique (CNRS) of France, and the University of Hawaii. This research used the facilities of the Canadian Astronomy Data Centre operated by the National Research Council of Canada with the support of the Canadian Space Agency. CFHTLenS data processing was made possible thanks to significant computing support from the NSERC Research Tools and Instruments grant program. Some of the results in this paper have been derived using the HEALPix~\cite{gorski05} package.

\appendix

\label{lastpage}

\end{document}